\documentstyle[aps,twocolumn,floats,epsf]{revtex}

\begin{document}  
\draft
\title{News from the Adriatico Research 
Conference on ``Superconductivity, Andreev 
Reflection, and Proximity Effect in Mesoscopic 
Structures''}                      
         
\author{D.~C.~Ralph and V.~Ambegaokar}  
\address{Laboratory of Atomic and Solid State 
Physics, Cornell University, Ithaca, NY 14853}
\maketitle
     
\begin{abstract}
The Adriatico Research Conference on 
``Superconductivity, Andreev Reflection, and 
Proximity Effect in Mesoscopic Structures'' took 
place at the International Center for Theoretical 
Physics in Trieste, Italy, July 8-11, 1997.  The 
organizers were Elias Burstein, Leonid Glazman, 
Teun Klapwijk, and Subodh R. Shenoy.  We 
describe some of the central issues discussed at the 
conference, along with more personal reflections 
prompted by new developments.
\end{abstract}     
\par                                         
\vspace{0.5cm}


\section{THE PROXIMITY EFFECT AND 
ANDREEV REFLECTION}

	Introductory treatments of the 
superconducting proximity effect~-- how electrons 
behave in the vicinity of an interface between a 
normal metal and a superconductor~-- typically 
follow one of two tracks.  One is to consider a 
Ginzburg-Landau picture of a 
superconductor/normal metal (SN) contact 
\cite{Deutscher}.  In this treatment, which is 
accurate for temperatures close to the 
superconducting critical temperature $T_c$, one 
may define a position-dependent electron-pair 
correlation function 
$<\Psi_{\downarrow}(x)\Psi_{\uparrow}(x)>$ 
which extends from the superconductor into the 
normal metal, decreasing exponentially (in diffusive 
materials) on thermal diffusion length scale of $L_T 
= \sqrt{\hbar D/ (k_{B}T)}$ (here $D$ is the 
diffusion constant and $T$ is the temperature).  This 
is perhaps a useful pedagogical approach, in that it 
allows one to think in a simple way that the Cooper 
pairs in the superconductor may ``leak'' through a 
non-zero thickness of normal metal.  However, this 
framework is of limited practical value for 
temperatures much lower than $T_c$, in that it does 
not provide a way to calculate experimental 
quantities such as supercurrents, the tunneling 
density of states in the normal metal, or the 
conductance properties for various geometries of 
superconductors and normal metals made into 
devices.  The problem is that the simple 
Ginzburg-Landau theory 
does not properly reflect the energy 
dependence of electronic properties.  In fact, in the 
low-temperature regime where the electron 
phase-breaking length is long enough to be ignored, the 
appropriate length scale for describing the range of 
pair-correlated electrons diffusing inside a normal 
metal that is in contact with a superconductor is 
given by the energy-dependent quantity 
$L_{\varepsilon} = \sqrt{\hbar D / \varepsilon}$, 
where $\varepsilon$ is the electronic energy 
measured with respect to the Fermi level.  This can 
be much longer than $L_T$.  A theory which 
properly takes this energy dependence into account 
is the ``quasiclassical Green's function theory'', 
formulated in general by Eilenberger 
\cite{Eilenberger}, and specialized to diffusive 
systems by Usadel \cite{Usadel} (some of their 
work being done during post-doctoral stays at our 
home institution -- Cornell!).  ``Quasiclassical'' 
means that the full, non-equilibrium Gorkov 
equations of superconductivity are coarse-grained 
so as to eliminate quantum features on the fine scale 
of $1/k_F$.  An overview on the current status of 
these methods was given at the conference by Gerd 
Sch\"{o}n.

	A second pedagogical approach to 
understand the proximity effect is Andreev 
reflection \cite{Andreev,BTK}.  In this picture one 
notes that if an electron is traveling from a normal 
metal into a superconductor, and it has an energy 
within a range of the superconducting gap $\Delta$ 
about the Fermi energy, then it cannot simply be 
transmitted into the superconductor because of the 
superconducting gap.  Instead, one may take the 
view that the electron, upon encountering the 
superconducting interface, produces a Cooper pair 
which is transmitted into the superconductor, and in 
the process a hole is retroreflected back into the 
normal state (in this way conserving charge, energy, 
and transverse momentum).  The physics of 
Andreev reflection is contained within the 
quasiclassical Green's function treatment.  However, 
in addition, more recent formulations have extended 
a purely scattering-theory approach to include 
coherent multiple processes in which both Andreev 
and normal reflections occur at imperfect SN 
interfaces, and electrons may also be scattered from 
defects within the metals.  The overall transport 
properties are than calculated in the spirit of the 
Landauer-B\"{u}ttiker formula, as a function of the 
transmission eigenvalues of an overall scattering 
matrix.  Colin Lambert reviewed the development 
of these methods, and Carlo Beenakker described 
related results which take into account statistical 
random-matrix properties of the scattering matrix.

	The central theme of the conference was a 
clear consensus that both flavors of theory, the 
quasiclassical Green's function method and the 
scattering matrix approach, are equivalent in the 
regimes in which they are both applicable, and that 
their main results are well-supported by recent 
experiments.  In the words of Michel Devoret, the 
proximity effect and Andreev reflection are ``two 
sides of the same coin.''  Nathan Argaman went so 
far as to show that the Usadel equations of the 
Green's function theory for diffusive metals may be 
derived within a purely scattering-matrix approach 
based on multiple Andreev scattering.  The two 
types of theories do have slightly different ranges of 
applicability.  Green's function methods begin to 
break down in applications to nm-scale devices with 
just a few conducting channels, a regime in which 
the scattering matrix methods are particularly 
suited.  Scattering matrix methods can also be used 
more easily to model sample-to-sample variations in 
the mesoscopic size regime.  However the Green's 
functions methods are otherwise more general, as 
they may include effects of a wide range of 
interactions, and also possibilities for the 
modification of superconducting regions due to 
their contact with normal metals, not 
included in present scattering-matrix treatments.  
The Green's functions are the sole means to 
calculate quantities such as magnetization, that 
cannot be related directly to transmission 
coefficients.

	The driving force for renewed interest in the 
proximity effect in the last 5 years is that new 
technologies for the fabrication of small electronic 
devices have allowed SN devices to be studied in 
size regimes that have never before been accessible 
-- smaller than both the low-temperature 
phase-breaking length for electrons ($L_{\phi}$), and also 
$L_T$.  These samples have been used to test the 
proximity-effect theories through measurements of 
densities of states, transport properties, and 
magnetization.  

	{\em Density of states:}  Michel Devoret 
reported tunneling measurements on small copper 
wires connected at one end to superconducting 
aluminum pads.  The density of states at different 
distances from the interface was probed with tunnel 
junctions a few 10's of nm wide.  The results 
displayed excellent agreement with calculations 
performed by Gerd Sch\"{o}n's group within the 
quasiclassical theory.  As predicted, the density of 
states in the normal metal was depressed for 
energies below an scale corresponding to $\hbar 
D/x^2$, where $x$ was the distance from the 
interface.  This is the energy range over which 
electrons diffusing a distance $x$ will all remain in 
phase.  For the best fit to theory, an effective 
spin-flip scattering time of approximately 65~ps was 
required.  The explanation of this somewhat 
unexpectedly short time is perhaps at this moment 
unclear.

	{\em Reentrant resistance:}  Measurements 
of the transport properties of well-characterized 
diffusive SN devices having a variety of different 
geometries were described by representatives of 
several groups, including M.~H.~Devoret, 
H.~Courtois, and B.~J.~van~Wees.  In accord with 
theory, the resistance of diffusive SN wires shows a 
``reentrance effect'' as a function of temperature, 
meaning that as the temperature is lowered below 
$T_c$, the resistance first decreases and then rises 
to begin to approach the normal state value as the 
temperature goes to zero.  Both quasiclassical 
theory and matrix scattering approaches predict that 
at $T=0$ the resistance of a disordered SN wire is 
precisely the normal state resistance, and the 
temperature scale for the minimum in the resistance 
is the Thouless energy $\hbar D / L^2$, divided by 
$k_B$ ($L$ is the length of the normal region of the 
wire).  

	{\em Interferometer devices:}  A clever trick 
employed by many groups (including V. Petrashov, 
M. H. Devoret, H. Courtois, and B. J. van Wees) is 
to make ``interferometer'' devices which consist of 
an open loop of superconductor whose ends are 
attached to different points of a normal metal 
conductor.  An applied magnetic field then acts to 
change the relative superconducting phase of the 
two ends, and allows phase-dependent 
measurements of Andreev-scattering effects.  The 
results of these experiments are all apparently in 
good qualitative accord with predictions.  In 
particular, for interferometer devices in which there 
is good metallic contact between the 
superconductors and the normal metals, the 
conductance oscillates periodically with the 
superconducting phase difference, with a relative 
amplitude which scales roughly as $\hbar D / (L^2 
k_B T)$, so that as a function of increasing 
temperature the oscillation amplitude falls slowly as 
$1/T$ rather than exponentially.  The interpretation 
of this result is straightforward, in that while 
electrons at the Fermi energy within an energy 
window of $k_{B}T$ contribute to the total 
conductance, only those within an energy window 
of $\hbar D / L^2$ remain in phase over the sample.

	{\em Disordered-enhanced Andreev 
reflection at tunnel barriers:}  Devices in which a 
superconductor and a normal metal are not in 
metallic contact, but are separated by a tunnel 
barrier, can also exhibit Andreev reflection 
processes.  For ballistic metal samples joined by 
tunnel barriers, this effect is predicted to be very 
weak \cite{BTK}.  However, disorder in the normal 
metal can enhance Andreev processes by orders of 
magnitude (a factor of 1000 for the Saclay group), 
due to an effect dubbed ``reflectionless tunneling''.  
The mechanism may be understood in analogy to 
the Fabry-Perot interferometer in optics.  In a 
disordered sample, an electron of energy 
$\varepsilon$ may be viewed as taking a path in 
which it undergoes many ordinary reflections from 
the SN tunnel junction and disorder, thereby 
returning to the tunnel junction many times.  At 
each reflection from the SN interface, there will be 
a small amplitude for Andreev reflection.  However, 
because Andreev reflection is a retro-reflection 
process, the reflected hole state (corresponding at 
$V=0$ to the electron energy $- \varepsilon$) will 
have almost precisely the same trajectory as the 
electron path (but in reverse), and the 
quantum-mechanical phase accumulated by 
the hole between 
reflections at the SN interface will match that of the 
electron.  The end result is that the amplitudes for 
Andreev reflection at each scattering event at the 
SN tunnel junction will add constructively, 
producing a much larger tunneling signal than if the 
processes were added incoherently.  With a voltage 
applied across the tunnel junction, the differences in 
energy (and hence wavelength) of the electron and 
reflected hole states will grow and the constructive 
interference will gradually be degraded, leaving a 
zero-bias peak in the conductance.  Another related 
effect, ``giant Andreev reflection'', was predicted by 
Beenakker and observed by van Wees for the 
geometry of a ballistic constriction in series with a 
disordered conductor.

	{\em Resistance increases due to 
superconductivity:}  Frank Wilhelm proposed a 
theory to describe the counterintuitive experimental 
result (Petrashov) that the resistance of a wide, 
diffusive normal metal wire in contact with a 
superconductor can increase as the sample is cooled 
through the superconductor's $T_c$.  The 
explanation appears to be a geometrical effect -- a 
consequence of the quasi-2-dimensional nature of a 
wide wire and the fact that current and voltage 
probes were positioned on opposite sides of the 
wire.

	{\em Supercurrents:}  Experiments on 
diffusive SNS devices exhibiting supercurrents were 
not discussed at the conference in as much depth as 
conductance measurements.  However, from the 
work of Courtois, it seems clear that supercurrents 
(at least in ``long'' devices where the N region is 
longer than the coherence length) are governed 
by the same energy scale, the Thouless energy $E_c 
= \hbar D/L^2$, which plays the central role in 
conductance measurements.  The typical magnitude 
of the critical current at $T=0$ is $I_c = E_c/R_n$, 
where $R_n$ is the normal state resistance of the 
device and the length scale in $E_c$ is the extent of 
the normal region.

	{\em Ballistic samples:}  Conductance 
measurements for 2-dimensional electron gas 
(2DEG) samples in which electron motion is 
ballistic, or quasiballistic (as opposed to diffusive) 
were presented by H. Takayanagi and A. F. 
Morpurgo.  Morpurgo described a breakdown of the 
idea of simple retroreflection of the hole in Andreev 
scattering, when the interface contains disorder on 
the scale of the electron wavelength.  Nevertheless, 
he argued that his experiments could still be 
described well by a semiclassical ray-tracing 
procedure in which the possible paths for electrons 
and holes were added coherently.  It appears that 
additional work is still required to test the behavior 
of even smaller and cleaner devices (such as those 
containing ballistic point contacts) where 
semiclassical ideas break down and a fully quantum 
picture is required.  It also seems that the 
characterization of ballistic 2DEG samples can be 
quite difficult, particularly concerning the quality of 
the coupling between the superconductor and the 
2DEG.  In present-generation devices, electron 
scattering is likely much stronger at this interface 
than in the 2DEG away from the superconductor.

	{\em Magnetization:}  Magnetization 
measurements appear to be an area which may 
provide a challenge for existing theory.  Joe Imry 
described measurements made by A. C. Mota 
(Zurich) on the susceptibility of fine 
superconducting wires (Nb) with a thin coating of 
normal metal (Cu or Ag).  As a function of 
decreasing temperature, the susceptibility becomes 
increasingly diamagnetic as the sample is cooled 
below the superconducting $T_c$, as is expected 
due to the proximity effect in the normal metal.  
However, when the samples are cooled even farther, 
to the low mK range and below, the susceptibility in 
some wires reaches an extremum and then turns 
around, so that in the $T=0$ limit some wires 
display a susceptibility that is even more 
paramagnetic than at the superconductor's $T_c$.  
Imry speculated that this behavior may be related to 
specific ``whispering gallery'' modes in the normal 
metal which may decouple from the superconductor 
well below $T_c$.

\section{SUPERCONDUCTING BREAK 
JUNCTIONS AND MULTIPLE ANDREEV 
REFLECTION}

	One of the most beautiful and interesting 
experiments discussed at the conference was the 
work of Elke Scheer and collaborators at Saclay, 
reported by Christian Urbina.  They were able to 
make detailed measurements of the current traveling 
via discrete quantum mechanical modes in 
atomic-scale superconducting (Al) break junctions.  This 
work may be viewed as a continuing development 
of the point-contact spectroscopy technique 
pioneered by Igor Yanson, and reviewed by him at 
the conference.  The beauty of the new Saclay work 
is that an analysis of the conductance for applied 
voltages less than the superconducting gap of the 
electrodes ({\em i.e.}, the subharmonic gap 
structure) was used to characterize the transmission 
coefficient for {\em each} of the active transport channels 
in the atomic-scale contact.  The theory of the 
subharmonic gap structure, developed by J. C. Cuevas and 
collaborators (Madrid) using a Hamiltonian approach, and
described by V. 
Shumeiko and D. V. Averin in a scattering-theory
picture, is remarkably in such 
good shape that the transmission coefficients of 6 or 
more different quantum channels may be 
determined simultaneously.  For aluminum break 
junctions, the Saclay group found, using fits
to the theory of the Madrid group, that at 
least three partially-transmitting channels were 
required to describe their data, no matter how small 
they made their contact.  Atomic-scale aluminum 
contacts are therefore different than 2DEG point 
contacts where transport may occur via a single, fully 
transmitting conductance channel.  Urbina (referencing
Cuevas {\em et al.})
speculated that the behavior of the aluminum 
contacts was rooted in chemistry -- each aluminum 
atom can contribute 3 combinations of hybridized s 
and p atomic orbitals which lead to transport 
channels in a wire.  Aluminum electrodes joined 
together by (purely s-like) gold atoms, on the other 
hand, can be reduced in size to the point that only a 
single quantum mode contributes to the 
conductance.

	In view of the fact that all the channels 
observed so far in the aluminum break junctions are 
only partially transmitting, it seems very puzzling 
that histograms of the values of the total 
conductance in these samples still show peaks near 
quantized values of conductance (integer multiples 
of $2 e^2/h$), corresponding to individual, 
fully-transmitting quantum channels.

	In a poster presentation, P. Dieleman 
{\em et al.\ }from T. M. Klapwijk's 
group described shot noise 
measurements of 
superconductor-insulator-superconductor 
tunnel junctions which display 
subharmonic gap structure due to multiple Andreev 
reflection.  At values of voltage corresponding to 
subharmonic peaks, they found an increase in the 
magnitude of the shot noise above the classical 
value, consistent with the view that in multiple 
Andreev reflection several electrons are transmitted 
simultaneously.

\section{SUPERCONDUCTIVITY IN nm-SCALE 
PARTICLES}

	One afternoon of the conference was given 
over to consideration of the nature of 
superconductivity in nm-scale metal particles, small 
enough that the discrete electrons-in-a-box 
energy-level spacing is comparable to the superconducting 
gap.  One of us (D.C.R) described experiments in 
which these discrete levels were measured in single 
aluminum nanoparticles by a tunneling technique, 
and the effects on the spectra of a variety of 
different forces and interactions, including 
superconducting pairing, were analyzed.  Andrei 
Zaikin reviewed theoretical results that as the level 
spacing in a metal sample is increased to approach 
the superconducting gap, the pairing parameter 
within BCS theory should be different for even and 
odd numbers of electrons.  It was not clear how this 
effect might be measured experimentally, for two 
reasons.  The first is that the even and odd pairing 
parameters may not be observables -- what is 
measured in a tunneling experiment are energy 
differences between states with even and odd 
numbers of electrons, so that it is not a trivial matter 
to separate the different pairing parameters.  Also, 
for nanoparticles in the size range where even and 
odd differences become interesting, the level 
spacing due to independent-electron quantum 
confinement becomes comparable to gaps due to 
superconductivity, and it is not clear how to 
separate these two effects.

	Jan von Delft described a model for how 
spin pair-breaking due to an applied magnetic field 
will affect the eigenstates in a small 
superconducting particle.  Von Delft found that the 
discreteness of the electronic spectrum may change 
the nature of the superconducting transition, 
compared to the theory of Clogston and 
Chandrasekhar (C\&C) \cite{CC} which describes 
well the transitions observed in thin film samples 
where the electronic spectrum is effectively a 
continuum.  In the C\&C theory, the tunneling 
threshold changes discontinuously at the transition 
field, but for a small particle this change may be 
continuous, if the transition from the 
superconducting state involves the flipping of just a 
single electron spin.

\section{MEAN FIELD THEORIES AND 
BEYOND}

	Igor Aleiner described recent work with 
Altshuler on the theory of a tunneling anomaly in a 
superconductor in a magnetic field above the 
paramagnetic limit.  The interesting result is that 
even though the mean field order parameter is zero 
in this regime, there is a singularity in the density of 
states due to fluctuations in the order parameter.

	A. F. Andreev reported on his published 
work in which broken gauge symmetry is taken as 
the central defining characteristic of 
superconductivity, requiring a modified form of 
statistical mechanics.  Some in the audience 
(including one of the authors, V.A., in closing 
remarks) expressed the view that ``broken 
symmetry'' is merely a mean-field treatment of the 
interacting system, and that residual effects such as 
those discussed by Aleiner (above) also contain 
important physics.

	In his closing remarks, V.A. also said that 
since the conference was in some ways in honor of 
Andreev, it might be worth noting that a 
superconducting quasiparticle is built up from 
repeated virtual Andreev scattering against the 
mean-field order parameter.  This can be seen by 
writing the electron propagator as 
\[G(\varepsilon,z)=\frac{z+\varepsilon}{z^2 - 
\varepsilon^2 - \mid{\Delta}\mid^2} = \frac{1}{z - 
\varepsilon - \Delta^{*}(z+\varepsilon)^{-
1}\Delta}.\]
The last term in the last denominator (the electron 
self-energy) shows a normal electron being 
converted into a normal hole via the Andreev 
process, which here cannot be energy conserving 
because there is no analog of the voltage across an 
interface.

\section{LOOKING TO THE FUTURE}

	Also presented at the conference were topics 
that look more to the future, in that mysteries 
remain which suggest avenues for future work.

	{\em Andreev processes involving localized 
states:}  Z. Ovadyahu described data from 
superconductor-insulator-normal metal tunnel 
junctions in which the barrier material was the 
Anderson insulator indium oxide, which contains a 
high density of localized electronic states.  
Conductance measurements showed unusual 
zero-bias signals, and also features at voltages well above 
the superconducting gap.  Igor Yanson noted that 
the above-gap features are similar to signals seen in 
metallic SN point contacts; however it seems to us 
that the details of the zero-bias signals are at least 
suggestive that something more interesting than 
pinholes may be at work.  Andreev reflection 
involving localized states could conceivably serve 
as an excellent model system for exploring the 
interplay of Coulomb charging effects, Kondo 
physics, and superconducting pairing.  Similar 
themes were touched upon by A. Golub in his talk, 
and R. Fazio and R. Raimondi in a poster.

	{\em Nonequilibrium effects:}  The study of 
nonequilibrium processes such as charge imbalance 
and phase slip centers has a proud history that was 
reviewed in a talk by M. Tinkham.  However, new 
non-equilibrium experiments in the mesoscopic 
regime continue to show unanticipated behavior, 
particularly when the samples are exposed to AC 
signals, as described by V. Chandrasekhar.

	{\em d-wave superconductors:}  Thus far all 
the proximity-effect devices that we have described 
utilized conventional s-wave superconductors.  Yu. 
S. Barach and Y. Tanaka provided a theoretical 
discussion of a variety of ways in which tunnel 
junctions made using high-$T_c$ d-wave 
superconductors would produce qualitatively 
different results.  These include an anomalous 
temperature dependence for the Josephson current, 
the existence of quasiparticle states bound to the 
tunnel barrier, and surface pair breaking.  Actually 
fabricating well-controlled high-$T_c$ tunnel 
junctions will be a daunting task because of their 
difficult chemistry, but O. Fischer showed that 
low-temperature STM studies of the superconducting 
cuprates are already providing interesting results.  
He demonstrated striking differences in tunneling 
spectra for the electronic states in the cores of 
magnetic vortices in high-$T_c$ materials, as 
compared to s-wave $NbSe_2$.  

	{\em Electron-electron interactions:}  The 
subject of electron-electron interactions in metals 
was not the focus of any scheduled presentations, 
but a recent paper by Mohanty, Jariwala, and Webb 
\cite{Mohanty} was the object of much 
informal debate.  This work proposes that 
zero-point fluctuations of the electromagnetic 
environment can produce electron dephasing in 
mesoscopic devices.  Michel Devoret also 
mentioned puzzling results out of Saclay, where 
direct measurements of electron energy relaxation 
processes have suggested a scaling form that is not 
compatible with a present understanding of 
interaction processes.

	Looking perhaps even farther into the future, 
F. Hekking reported calculations of the properties of 
superconductors coupled to the interacting electrons 
in one-dimensional Luttinger liquids, and H. Mooij 
speculated as to the use of Josephson-junction 
devices for quantum computations.  There is no way 
to know at this point the prospects for whether the 
quantum coherence of Josephson junctions can be 
controlled sufficiently to allow for real quantum 
computations, but we expect that the 
macroscopic-quantum physics to be learned in 
this effort will be 
fascinating.  An important intermediate goal on the 
way to computation will be to attain sufficiently 
long coherence times to produce a quantum clock 
using Josephson junctions, something long sought 
but without success to date.

\section{PERSONAL REFLECTIONS}

	Perhaps one way to summarize the present 
status of the theory of the superconducting 
proximity effect might be to paraphrase a remark 
made by Gerd Sch\"{o}n concerning the 
quasiclassical theory, ``You give the theory to 
students, they solve some differential equations, and 
before long they come back with results!''  Good 
agreement between recent experiments and theory 
give considerable confidence that a reasonably 
comprehensive understanding of (s-wave) 
superconducting/normal metal interfaces is close at 
hand.  Before embarking upon triumphalism, 
however, we note that while the Green's function 
theory needed to explain most of the recent 
generation of proximity-effect experiments was 
complete long before the experiments were begun, 
there was still a delay of some years after the first 
experiments before their explanation was generally 
appreciated.  Part of the difficulty undoubtedly lay 
in uncertainty about experimental parameters 
(especially the quality of the interfaces between the 
superconductors and normal metals), but we also 
believe that there continue to be important issues of 
accessibility in the theory.  We suggest that one of 
our goals, as a field of study,  should be the further 
development of reliable tools for working intuition, 
so that those who live happy lives without benefit of 
Green's functions may have good pictures with 
which to begin to understand 
superconducting/normal metal devices, and a clear 
prescription for how to proceed in reliable 
modeling.  We need popularizers, not prophets.

	Important steps along these lines were 
reported at the conference.  Yuli Nazarov's ``circuit 
theory'' formulation for the quasiclassical Green's 
functions is a valuable contribution, though it still 
cannot be said to be optimally ``user friendly''.  We 
find the scattering-matrix theories of Andreev 
reflection in SN devices to be very important as a 
more intuitive approach than the Green's function 
theory in many situations.  Nathan Argaman's poster 
was particularly interesting in this regard, as it 
showed explicitly that the main formulas describing 
the proximity effect in the quasiclassical Green's 
function theory can in fact be derived from a simple 
picture involving nothing more than multiple 
Andreev reflection.  In addition, we appreciated the 
strategy that Bart van Wees took in his talk to help 
build intuition.  In the spirit of Nazarov's circuit 
theory, he considered the nature of Andreev 
scattering in the important cases of a tunnel barrier, 
a disordered wire, and a ballistic constriction, and 
then he considered what happens when these 
elements are combined.

\vspace{-0.3cm}

\section{ACKNOWLEDGMENTS}

In addition to the sponsorship of the International 
Center for Theoretical Physics (ICTP), this 
Adriatico Research Conference was aided by 
support funds from the United States Air Force 
European Office of Aerospace Research and 
Development.  The authors' work is supported by 
the US NSF (Grants DMR-9407245, 
DMR-9632275, DMR-9705059), 
ONR (Grant N00014-97-1-0745), and the 
A. P. Sloan Foundation.

\end{document}